\documentclass[iop,apjl]{emulateapj}

\usepackage{graphicx}  % needed for figures
\usepackage{dcolumn}   % needed for some tables
\usepackage{bm}        % for math
\usepackage{amsfonts,amsmath,amssymb,mathrsfs}
\usepackage{color}
\usepackage{epsfig}
\usepackage{natbib,times}
\usepackage{url}
\usepackage{times}
\allowdisplaybreaks[4]

\pdfoutput=1

\makeatletter
\newcommand{\vast}{\bBigg@{4}}
\newcommand{\Vast}{\bBigg@{5}}
\makeatother
\DeclareMathAlphabet{\mathpzc}{OT1}{pzc}{m}{it}

\shorttitle{General relativistic plasma with Landau-Lifshitz radiation reaction}
\shortauthors{Wenshuai Liu}

\begin{document}

\title{One-fluid equations of general relativistic two-fluid plasma with the Landau-Lifshitz radiation reaction force in curved space}

\author{Wenshuai {Liu}\altaffilmark{1}$^{,a}$, Weihao {Bian}\altaffilmark{1}$^{,b}$, Bixuan {Zhao}\altaffilmark{1}, Liming {Yu}\altaffilmark{1}, Chan {Wang}\altaffilmark{1}}

\altaffiltext{1}{School of Physics and Technology, Nanjing Normal University, Nanjing 210023, China}
\altaffiltext{a}{674602871@qq.com}
\altaffiltext{b}{whbian@njnu.edu.cn}

\begin{abstract}
 Incorporating the radiation reaction force into two-fluid plasma in curved space, we get a set of one-fluid general relativistic magnetohydrodynamics (GRMHD) equations with the Landau-Lifshitz radiation reaction force. We analyze the importance of the radiation reaction acting on plasma around an astrophysical compact object.
\end{abstract}


%%--------------------------------------------------%%
\section{Introduction}
A charged particle moving in the electromagnetic field can feel Landau-Lifshitz radiation reaction force due to synchrotron radiation, thus modifying the motion of the charged particle significantly when taking into account the radiation reaction force. Recent observations have demonstrated that there is strong evidence that a magnetic field of several hundred gauss exists in the vicinity of the supermassive black hole at the center of the Milky Way \citep{1}. A dynamo mechanism from an accretion disk around a black hole accounts for the appearance of such a magnetic field \citep{2,27,28}. There is also clear evidence that the magnetic field on the surface of the neutron star can be up to $10^{14}$ G \citep{21,22,23,24,25,26}. The Landau-Lifshitz radiation reaction has been investigated in detail in \cite{15} in flat space, while, in curved space, the radiation reaction is described in \cite{770}, \cite{771}, \cite{772} and \cite{49}. Thus, the radiation reaction force can affect the charged particles from the accretion disk around the black hole and the neutron star significantly. Radiation reactions have been an important element of pair creation scenarios in positron-electron plasma just above the pole of the event horizon \citep{b22}. Curvature radiation in black hole magnetopshere pair creation schemes is radiation resistance limited \citep{b11}. Recently, a radiation reaction has also been applied to protonic acceleration in the vortex above the pole of the black hole \citep{b33}. Radiation reactions are of fundamental importance in the evacuated vortex of black hole magnetospheres \citep{2}. In particular, if the field line angular velocity is set much less than the horizon angular velocity by distant plasma and a very tenuous plasma exists in the event horizon magnetosphere then radiation resistance will determine the flow dynamics of accretion as well as the rotational energy extraction by a putative jet \citep{2,b44}. The dynamics mentioned above cannot be revealed by MHD simulations with mass floors. The ad hoc injection of mass will damp any large waves that can break ideal MHD and prevent the associated large local electromagnetic forces from being achieved. Therefore, all existing numerical simulations of the black hole magnetosphere bypass the radiation reaction dominated dynamics as a consequence of numerical dissipation of waves and numerical diffusion in the MHD system in the evacuated vortex above the event horizon. Thus, a proper treatment of radiation reaction is critical for assessing the time evolution of these types of astrophysical systems. Numerical simulations using general relativistic magnetohydrodynamics (GRMHDs) are applied to investigate the physical process in accretion disks around neutron stars, as well as in microquasars ($\mu$QSOs), gamma-ray bursts and active galactic nuclei. In curved spacetime, the dynamical evolution of the high energy disks made of ion-electron plasma in simulated GRMHD is usually performed without the contribution from Landau-Lifshitz radiation reaction force which may play a crucial role. The method of particle-in-cell \citep{4,7,6,8,10,9} containing the radiation reaction forces has been investigated in flat space, while, in curved space, the same method with a radiation reaction has been achieved in \cite{5}. Incorporating the radiation reaction into the relativistic magnetohydrodynamic equations governing the dynamics of plasma has been studied by \cite{11}, \cite{12} and \cite{13}. In a recent study, \cite{773} achieved the one-fluid relativistic magnetohydrodynamics description of two-fluid plasma in which the Landau-Lifshitz radiation reaction is incorporated. However, numerical simulation with the radiation reaction from \cite{773} is not practical due to its highly nonlinear form at present, we expect that analytical investigations are essential and provide more motivation for future work. Thus, in this work, similar to \cite{773}, we get the GRMHD equation for the one-fluid description of two-fluid plasma containing a Landau-Lifshitz radiation reaction in curved space, and these results could be applied to both positron-electron and proton-electron plasma.

\section{Equations derived}
In this section, we derive the one-fluid GRMHD including the Landau-Lifshitz radiation reaction force based on general relativistic two-fluid plasma in curved space. The two-fluid plasma is composed of positively charged particles with mass $m_+$, electric charge $e$ and negatively charged particles with mass $m_-$, electric charge $-e$. The spacetime is $(t,x^1,x^2,x^3)$ where a line element is $ds^2=g_{\mu\nu}dx^\mu dx^\nu$ characterized by metric $g_{\mu\nu}$. We set $c=1$, $\epsilon_0=1$, and $\mu_0=1$ which represent the speed of light, the dielectric constant, and the magnetic permeability in vacuum set to be unity.

Before deriving the one-fluid GRMHD equations based two-fluid plasma with the radiation reaction, we first define the average and difference variables which are the same as those of \cite{2010}. We list the variables as follows
\begin{eqnarray}
\rho &=& m_+ n_+ \gamma_+' + m_- n_- \gamma_-' \\
n &=& \frac{\rho}{m} \\
p &=& p_+ + p_- \\
\Delta p &=& p_+ - p_- \\
U^\mu &=& \frac{1}{\rho} ( m_+ n_+ u_+^\mu + m_- n_- u_-^\mu )
\label{1} \\
J^\mu &=& e(n_+ u_+^\mu - n_- u_-^\mu)
\label{2}
\end{eqnarray}

where the variables with subscripts, plus and minus, are those of the fluid of particles with positive charge and of the fluid of particles with negative charge, respectively,
$n_\pm$ is the proper particle number density, $p_\pm$ is the proper pressure, $\gamma_\pm'$ is the Lorentz factor of the positively charged and negatively charged fluid observed by the plasma's local center-of mass frame, $m=m_+ + m_-$, $u_\pm^\mu$ is the four-velocity, and $\rho$, $n$, $p$, $U^\mu$ and $J^\mu$ are density, proper particle number density, proper pressure, four-velocity, four-current density in one-fluid frame, respectively.

The generalized GRMHD equations based on the two-fluid plasma with the Maxwell equations are given
as follows \citep{2010}
\begin{eqnarray}
\nabla_\nu (\rho U^\nu) &=& 0 , \label{gg7}\\
\nabla_\nu  \left [
h \left (U^\mu U^\nu + \frac{\mu}{(ne)^2} J^\mu J^\nu \right ) \right ]
&=& -\nabla^\mu p + J^\nu {F^\mu}_\nu ,  \label{gg8}\\
 \frac{1}{ne}  \nabla_\nu \left [ \frac{\mu h}{ne} \left \{
U^\mu J^\nu + J^\mu U^\nu
- \frac{\Delta \mu}{ne} J^\mu J^\nu \right \} \right ]
&=& \frac{1}{2ne} \nabla^\mu (\Delta \mu p - \Delta p) +
\left ( U^\nu - \frac{\Delta \mu}{ne} J^\nu \right) {F^\mu}_\nu
- \eta [J^\mu - \rho_{\rm e}' (1+\Theta) U^\mu] \\ \label{gg9}
 \nabla_\nu \hspace{0.3em} ^*F^{\mu\nu} &=& 0 \\ \label{gg10}
%% \label{frl4f} \\
 \nabla_\nu F^{\mu\nu} &=& J^\mu  \label{gg11}
%% \label{aml4f}
\end{eqnarray}
where $\mu = m_+ m_-/m^2$, $\Delta \mu = (m_+ -m_-)/m$, $q=ne$, $Q=U^\nu J_\nu$, $F_{\mu\nu}$ is the electromagnetic field tensor, $\hspace{0.3em} ^*F^{\mu\nu}$ is the dual tensor density of $F_{\mu\nu}$, $\mu, \nu=0,1,2,3$ and $\Theta$ is the thermal
energy exchange rate from the negatively charged fluid to the positive fluid (see Appendix A of \cite{14} for details).

In Minkowski spacetime, the dynamical equation of a charged particle with mass $m$ and electric charge $e$ when accounting for Landau-Lifshitz radiation reaction is
\begin{equation}
m\frac{du^{\mu }}{ds}=eF^{\mu \nu }u_{\nu }+g^{\mu }
\label{eq01a}
\end{equation}
where $u_{\mu }$ is the four-velocity and $\mu =0,1,2,3$

The radiation reaction force in the Lorentz-Abraham-Dirac form is

 \begin{equation}
g^{\mu }=\frac{2e^{2}}{3}\left[ \frac{d^{2}u^{\mu }}{ds^{2}}-u^{\mu }u^{\nu }\frac{d^{2}u_{\nu }}{ds^{2}} \right]
\label{eq02}
\end{equation}

When assuming that $g^\mu$ is small compared with Lorentz force in the instantaneous rest frame of the charged particle \citep{15}, we could express equation (\ref{eq02}) as follows
\begin{eqnarray}
g^{\mu }&=&\frac{2e^{3}}{3m}\left\{ \frac{\partial F^{\mu \nu }}{
\partial x^{\lambda }}u_{\nu }u_{\lambda }
-\frac{e}{m}\left[ F^{\mu
\lambda }F_{\nu \lambda }u^{\nu }-\left( F_{\nu \lambda }u^{\lambda }\right)
\left( F^{\nu \kappa }u_{\kappa }\right) u^{\mu }\right] \right\}\label{eq10}
\end{eqnarray}
where the first term, called Frenkel force \citep{20}, is negligible compared to other terms \citep{9,16}. Then equation (\ref{eq01a}) becomes
\begin{eqnarray}
m\frac{du^{\mu }}{ds} & = & eF^{\mu \nu }u_{\nu }
-\frac{2e^{4}}{3m^{2}}\left[ F^{\mu
 \lambda }F_{\nu \lambda }u^{\nu }-\left( F_{\nu \lambda }u^{\lambda }\right)
 \left( F^{\nu \kappa }u_{\kappa }\right) u^{\mu }\right]
\end{eqnarray}

The above derived radiation reaction force acting on a charged particle is in Minkowski spacetime. When changing to curved spacetime, we next derive the detailed form of the radiation reaction felt by a charged particle. Without the radiation reaction, the dynamical equation of a charged particle with mass $m$ and electric charge $q$ is
\begin{equation}
   \frac{D u^\mu}{d \tau}= \frac{q}{m} F^{\mu\nu} u_{\nu}
   \label{cureqmogen1}
\end{equation}

The dynamical motion of a charged particle undergoing radiation reaction force in curved space is \citep{772,774,49}
\begin{equation}
 \frac{D u^\mu}{d \tau} = \frac{q}{m} F^{\mu\nu} u_{\nu}
+ \frac{2 q^2}{3 m} \left( \frac{D^2 u^\mu}{d\tau^2} -u^\mu u^\nu \frac{D^2 u_\nu}{d\tau^2} \right)
 + \frac{q^2}{3 m} \left(R^{\mu}_{\,\,\,\lambda} u^{\lambda} + R_{\nu\lambda} u^{\nu} u^{\lambda} u^{\mu} \right) + \frac{2 q^2}{m} ~f^{\mu \nu}_{\rm \, tail} \,\, u_\nu,
\label{eqmoDWBH}
\end{equation}
where
\begin{equation}
f^{\mu \nu}_{\rm \, tail}  = \int_{-\infty}^{\tau-0^+}
D^{[\mu} G^{\nu]}_{ + \lambda'} \bigl(z(\tau),z(\tau')\bigr)
u^{\lambda'} \, d\tau'
\label{vv}
\end{equation}
is the tail integral.

The term with Ricci tensor vanishes in the metric, being not included in the derivation \citep{49}. The tail term is the integral of the past world line of the particle, which can be neglected as we show in the following. For a charged particle with mass $m$ and electric charge $q$, the tail term $\sim G M q^2/(r^3 c^2)$ while the Newton gravitational force $\sim  G M m/r^2$, then we get the ratio of the tail term to the Newton gravitational force near a black hole's horizon ($r\sim 2 G M/c^2$) as \citep{49}

\begin{equation}
         \frac{F_{\rm tail}}{F_{\rm N}}   \sim     \frac{q^2}{m M G}   \sim   10^{-19} \left(\frac{q}{e}\right)^2 \left(\frac{m_e}{m}\right) \left(\frac{10 M_{\odot}}{M}\right),
\end{equation}
where $m_e$ and $e$ are the mass and the charge of an electron

The ratio is eight orders lower when taking supermassive black hole with mass $M \sim 10^9 M_{\odot}$ \citep{49}. The second term from the right-hand side of equation (\ref{eqmoDWBH}) $\sim q^4 B^2/(m^2 c^4)$ when velocity is comparable to the speed of light, then the ratio of this term to Newton gravitational force is \citep{49}
\begin{equation}
 \frac{F_{\rm RR}}{F_{\rm N}}   \sim \frac{q^4 B^2 M G}{m^3 c^8}  \sim  10^{3} \left(\frac{q}{e}\right)^4 \left(\frac{m_e}{m}\right)^3 \left( \frac{B}{10^8 {\rm G}}\right)^2 \left(\frac{M}{10 M_{\odot}}\right).
\end{equation}
Considering the supermassive black hole with mass $M \sim 10^9 M_{\odot}$ and magnetic field $B \sim 10^4 {\rm G}$, the ratio leads to the same order of magnitude \citep{49}, meaning that the second term can be comparable to the gravitational force.

With the negligence of the third and the last terms on the right-hand side of equation (\ref{eqmoDWBH}), the equation reduces to
\begin{equation}
 \frac{D u^\mu}{d \tau} = \frac{q}{m} F^{\mu\nu} u_{\nu}
+ \frac{2 q^2}{3 m} \left( \frac{D^2 u^\mu}{d\tau^2} - u^\mu u^\nu \frac{D^2 u_\nu}{d\tau^2} \right)
\label{gg1}
\end{equation}

Similar to the method in the special relativistic case, we take the covariant derivative with respect to the proper time from both sides of equation (\ref{cureqmogen1}) and get
\begin{equation}
\frac{D^2 u^{\alpha}}{d\tau^2} = \frac{q}{m} \frac{D F^{\alpha\beta}}{d x^{\mu}} u^\beta u^\mu + \frac{q^2}{m^2} F^{\alpha\beta}
F_{\beta\mu} u^\mu,
\label{gg2}
\end{equation}

Then we get the radiation reaction force by substituting equation (\ref{gg2}) into (\ref{gg1})
\begin{equation}
\frac{D u^\mu}{d \tau} = \frac{q}{m} F^{\mu\nu} u_{\nu}+\frac{2q^2}{3m}  \left(\frac{q}{m}\frac{D F^{\alpha\beta}}{d x^{\mu}} u_\beta u^\mu - \frac{q^2}{m^2} \left( F^{\mu\alpha}
F_{\alpha\beta}u^\beta -  F^{\sigma\alpha} u_\alpha F_{\sigma\nu}  u^\nu u^\mu \right) \right),
\label{gg33}
\end{equation}
where
\begin{equation}
\frac{D F^{\alpha\beta}}{d x^{\mu}} = \frac{\partial F^{\alpha\beta}}{\partial x^{\mu}} + \Gamma^{\alpha}_{\lambda\mu} F^{\lambda\beta} +
\Gamma^{\beta}_{\lambda\mu} F^{\alpha\lambda}
\end{equation}
As in \cite{773}, $\frac{D F^{\alpha}_{\,\,\,\beta}}{d x^{\mu}} u^\beta u^\mu$ is negligible compared to other terms. Then equation (\ref{gg33}) changes to
\begin{equation}
m \frac{D u^\mu}{d \tau} = q F^{\mu\nu} u_{\nu}-\frac{2q^4}{3m^2}  \left( F^{\mu\alpha}
F_{\alpha\beta}u^\beta -  F^{\sigma\alpha} u_\alpha F_{\sigma\nu}  u^\nu u^\mu \right),
\label{gg4}
\end{equation}

When inserting the radiation reaction into generalized GRMHD, we need to get the total radiation reaction force on the two species of the two-fluid plasma in curved space. Using a similar method to that in \cite{773}, we should get $u^{\mu}_\pm$ and $n_\pm$. Before doing this, we first show the metric $g_{\mu\nu}$ \citep{2010}
\begin{equation}
g_{00}=-h_0^2, \verb!   !
g_{ii}=h_i^2,  \verb!   !
g_{i0}=g_{0i} =-h_i^2 \omega _i    ,
\label{defmt}
\end{equation}
where we assume that off-diagonal spatial elements vanish.

Then we get
\begin{equation}
ds^2 =  g_{\mu \nu} dx^{\mu} dx^{\nu} =-h_0^2 dt^2
  +\sum _{i=1}^3 \left [h_i^2(dx^i)^2 - 2h_i^2 \omega _i dt dx ^i
\right]   .
\label{defle}
\end{equation}

When defining
\begin{eqnarray}
\alpha = \left [ h_0^2+\sum _{i=1}^3
\left ( h_i \omega _i \right ) ^2 \right ]^{1/2} , \verb!  !
\label{defal}
%% \nonumber \\
\beta ^ {i} = \frac{h_i \omega _i}{\alpha }   ,
\label{diff_alpbet}
\label{defbe}
\end{eqnarray}
as a lapse function and shift vector, respectively, we get the line element as

\begin{equation}
ds^2=-\alpha ^2 dt^2+\sum _{i=1}^3 (h_i dx^i - \alpha \beta ^i dt)^2 .
\label{redle}
\end{equation}

The determinant of the matrix $g_{\mu\nu}$ is $ g \equiv - (\alpha h_1 h_2 h_3)^2$ and the the contravariant metric is given as
\begin{equation}
g^{00}=- \frac{1}{\alpha ^2}  , \verb!   !
g^{i0}=g^{0i}= - \frac{\omega_i}{\alpha^2} =- \frac{\beta^i}{\alpha h_i}, \verb!   !
g^{ij} = \frac{1}{h_i h_j} ( \delta ^{ij}
-\beta ^i \beta ^j ),
\label{defmtc}
\end{equation}
where $\delta ^{ij}$ is the Kronecker symbol.

When introducing a locally nonrotating frame, the zero-angular-momentum-observer (ZAMO) frame, it is convenient to write (\ref{defle}) as
\begin{equation}
ds^2= - d \hat{t}^2 + \sum _i (d \hat{x}^i)^2
= \eta_{\mu\nu} d\hat{x}^\mu d\hat{x}^\nu  ,
\label{redlez}
\end{equation}
where
\begin{eqnarray}
d\hat{t} = \alpha dt,
\label{transf1}
\\
d\hat{x}^i = h_i dx^i - \alpha \beta ^i dt.
\label{transf2}
\end{eqnarray}

Thus the spacetime is similar to Minkowski spacetime locally in the ZAMO frame. For contravariant vector$a^\mu$ in the Boyer¨CLindquist coordinates, the contravariant vector $\hat{a}^\mu$, which is in the ZAMO frame, is
\begin{equation}
\hat{a}^0 = \alpha a^0, \verb!   !
\hat{a}^i = h_i a^i - \alpha \beta ^i a^0 .
\label{fidocon}
\end{equation}

Then the covariant vector $\hat{a}_\mu$ is
\begin{equation}
\hat{a}_0 = \frac{1}{\alpha} a_0 + \sum _i \frac{\beta ^i}{h_i} a_i, \verb!   !
\hat{a}_i = \frac{1}{h_i} a_i .
\label{fidocov}
\end{equation}

Based on the above, we get
\begin{eqnarray}
\gamma &=& \hat{U}^0 = \alpha U^0 , \label{deg} \\
\hat{v}^i & \equiv & \frac{\hat{U}^i}{\hat{U}^0}
=\frac{h_i}{\gamma} U^i- \alpha \beta ^i \frac{U^0}{\gamma}
\end{eqnarray}
and we note that
\begin{equation}
\gamma = \frac{1}{\root \of {1-{\sum}_{i=1}^3 (\hat{v}^i)^2}}
\end{equation}

With respect to the two-fluid plasma, the Lorentz factor for the positively charged and negatively charged component are
\begin{equation}
\gamma_\pm = \frac{1}{\root \of {1-{\sum}_{i=1}^3 (\hat{v_\pm}^i)^2}}
\label{gg5}
\end{equation}

From equation (\ref{1}) and (\ref{2}), we get
\begin{equation}
n_\pm u_\pm ^\mu = \frac{1}{m} \left  (
\rho U^\mu \pm \frac{m_\mp}{e} J^\mu \right )
\label{gg6}
\end{equation}
Then we could get $u^{\mu}_\pm$ and $n_\pm$ as follows with equation (\ref{gg6}) and (\ref{gg5})
\begin{eqnarray}
n_\pm & = & \frac{\sqrt{[\alpha\left  (\rho U^0 \pm \frac{m_\mp}{e} J^0 \right )]^2- \sum_{n=1}^3[h_n\left(\rho U^n \pm \frac{m_\mp}{e} J^n \right )-\alpha\beta^n(\rho U^0 \pm \frac{m_\mp}{e} J^0)]^2}}{m}\label{17}\\
u^\mu_\pm & = & \frac{\left  (\rho U^\mu \pm \frac{m_\mp}{e} J^\mu \right )}{\sqrt{[\alpha\left  (\rho U^0 \pm \frac{m_\mp}{e} J^0 \right )]^2- \sum_{n=1}^3[h_n\left(\rho U^n \pm \frac{m_\mp}{e} J^n \right )-\alpha\beta^n(\rho U^0 \pm \frac{m_\mp}{e} J^0)]^2}}\label{18}
\end{eqnarray}

Then the radiation reaction force of the two species in the fluid element is
\begin{equation}
F^{\mu}_{LL\pm}=-\frac{2n_\pm e^{4}}{3m^{2}_\pm}\left[ F^{\mu
 \lambda }F_{\nu \lambda }u^{\nu }_\pm-\left( F_{\nu \lambda }u^{\lambda }_\pm\right)
 \left( F^{\nu \kappa }u_{\pm \kappa }\right) u^{\mu }_\pm\right]
 \label{eq9}
\end{equation}

Inserting the fluid element's radiation reaction equation (\ref{eq9}) into momentum density equation (\ref{gg8}), we get the momentum density equation with the radiation reaction as
\begin{eqnarray}
\nabla_\nu  \left [
h \left (U^\mu U^\nu + \frac{\mu}{(ne)^2} J^\mu J^\nu \right ) \right ]
&=& -\nabla^\mu p + J^\nu {F^\mu}_\nu
 \nonumber \\
& - & \frac{2n_+ e^{4}}{3m^{2}_+}\left[ F^{\mu
 \lambda }F_{\nu \lambda }u^{\nu }_+
 - \left( F_{\nu \lambda }u^{\lambda }_+\right)
 \left( F^{\nu \kappa }u_{+ \kappa }\right) u^{\mu }_+\right]
 \nonumber \\
& - & \frac{2n_- e^{4}}{3m^{2}_-}\left[ F^{\mu
 \lambda }F_{\nu \lambda }u^{\nu }_-
 - \left( F_{\nu \lambda }u^{\lambda }_-\right)
 \left( F^{\nu \kappa }u_{- \kappa }\right) u^{\mu }_-\right]
 \label{gg28}
\end{eqnarray}

Then equation (\ref{gg7})-(\ref{gg11}) can be changed to
\begin{eqnarray}
\nabla_\nu (\rho U^\nu) &=& 0 , \label{gg12}\\
\nabla_\nu  \left [
h \left (U^\mu U^\nu + \frac{\mu}{(ne)^2} J^\mu J^\nu \right ) \right ]
&=& -\nabla^\mu p + J^\nu {F^\mu}_\nu
 \nonumber \\
& - & \frac{2n_+ e^{4}}{3m^{2}_+}\left[ F^{\mu
 \lambda }F_{\nu \lambda }u^{\nu }_+
 - \left( F_{\nu \lambda }u^{\lambda }_+\right)
 \left( F^{\nu \kappa }u_{+ \kappa }\right) u^{\mu }_+\right]
 \nonumber \\
& - & \frac{2n_- e^{4}}{3m^{2}_-}\left[ F^{\mu
 \lambda }F_{\nu \lambda }u^{\nu }_-
 - \left( F_{\nu \lambda }u^{\lambda }_-\right)
 \left( F^{\nu \kappa }u_{- \kappa }\right) u^{\mu }_-\right]  \label{gg13}\\
 \frac{1}{ne}  \nabla_\nu \left [ \frac{\mu h}{ne} \left \{
U^\mu J^\nu + J^\mu U^\nu
- \frac{\Delta \mu}{ne} J^\mu J^\nu \right \} \right ]
&=& \frac{1}{2ne} \nabla^\mu (\Delta \mu p - \Delta p) +
\left ( U^\nu - \frac{\Delta \mu}{ne} J^\nu \right) {F^\mu}_\nu
- \eta [J^\mu - \rho_{\rm e}' (1+\Theta) U^\mu] \\ \label{gg14}
 \nabla_\nu \hspace{0.3em} ^*F^{\mu\nu} &=& 0 \\ \label{gg15}
%% \label{frl4f} \\
 \nabla_\nu F^{\mu\nu} &=& J^\mu  \label{gg16}
%% \label{aml4f}
\end{eqnarray}

For a convenient form, equations (\ref{gg12})-(\ref{gg16}) can be written as
\begin{eqnarray}
%% \nabla _\nu (\rho U^\nu) =
\frac{1}{\sqrt{-g}} \frac{\partial}{\partial x^\nu}
\left ( \sqrt{-g} \rho U^\nu \right )
&=& 0,
\label{gg17}
\\
%% \nabla _\nu T^{\mu \nu} =
\frac{1}{\sqrt{-g}} \frac{\partial}{\partial x^\nu}
\left ( \sqrt{-g} T^{\mu \nu} \right )
+\Gamma_{\sigma \nu}^\mu T^{\sigma \nu}
&=&  -  \frac{2n_+ e^{4}}{3m^{2}_+}\left[ F^{\mu
 \lambda }F_{\nu \lambda }u^{\nu }_+
 - \left( F_{\nu \lambda }u^{\lambda }_+\right)
 \left( F^{\nu \kappa }u_{+ \kappa }\right) u^{\mu }_+\right]
 \nonumber \\
& - & \frac{2n_- e^{4}}{3m^{2}_-}\left[ F^{\mu
 \lambda }F_{\nu \lambda }u^{\nu }_-
 - \left( F_{\nu \lambda }u^{\lambda }_-\right)
 \left( F^{\nu \kappa }u_{- \kappa }\right) u^{\mu }_-\right],
\label{gg18}
\\
\frac{1}{ne} \nabla_\nu K^{\mu\nu}
&=& \frac{1}{2ne} \nabla^\mu (\Delta \mu p - \Delta p)
+ \left ( U^\nu - \frac{\Delta \mu}{ne} J^\nu \right ) {F^\mu}_\nu
-\eta [J^\mu - \rho_{\rm e}'(1+\Theta) U^\mu]  .
\\
\partial _\mu F_{\nu \lambda} +
\partial _\nu F_{\lambda \mu} +
\partial _\lambda F_{\mu \nu}& =& 0   ,
\label{gg19}
\\
%% \nabla _\mu F^{\mu \nu} =
\frac{1}{\sqrt{-g}} \frac{\partial}{\partial x^\nu}
\left ( \sqrt{-g} F^{\mu \nu} \right ) &=&- J^\nu   ,
\label{gg20}
\end{eqnarray}
where
\begin{equation}
K^{\mu\nu} \equiv \frac{\mu h^\ddagger}{ne} \left (
U^\mu J^\nu + J^\mu U^\nu - \frac{\Delta h^\sharp}{neh^\ddagger} J^\mu J^\nu
\right ) + \frac{\Delta h}{2} U^\mu U^\nu,
\end{equation}

In the ZAMO frame, we could get the the following set of equations based on equation (\ref{gg17})-(\ref{gg20}) combined with radiation reaction force \citep{2010}

\begin{equation}
\frac{\partial \gamma \rho}{\partial t} = - \frac{1}{h_1 h_2 h_3} \sum _j
\frac{\partial}{\partial x^j} \left [
\frac{\alpha h_1 h_2 h_3}{h_j} \gamma \rho (\hat{v}^j + \beta^j)
\right ]   ,
\label{cmma}
\end{equation}
\begin{eqnarray}
\frac{\partial \hat{P}^i}{\partial t} &=& - \frac{1}{h_1 h_2 h_3} \sum _j
\frac{\partial}{\partial x^j} \left [
\frac{\alpha h_1 h_2 h_3}{h_j} (\hat{T}^{ij} + \beta ^j \hat{P}^i)
\right ]
- ( \epsilon + \gamma \rho) \frac{1}{h_i}
\frac{\partial \alpha}{\partial x^i}
\nonumber \\
&+& \sum_j \alpha \left [G_{ij} \hat{T}^{ij} - G_{ji} \hat{T}^{jj}
+ \beta^j (G_{ij} \hat{P}^i -G_{ji} \hat{P}^j)
\right ]
- \sum _j \sigma _{ji} \hat{P}^j +\alpha\hat{F^i_{LL}}  ,
\label{cmmo}
\end{eqnarray}
\begin{eqnarray}
\frac{\partial \epsilon}{\partial t} &=&  - \frac{1}{h_1 h_2 h_3} \sum _j
\frac{\partial}{\partial x^j} \left [
\frac{\alpha h_1 h_2 h_3}{h_j} (\hat{P}^{j} -\gamma \rho \hat{v}^j
+ \beta ^j \epsilon)
\right ]
\nonumber \\
&-& \sum _j \hat{P}^j \frac{1}{h_j} \frac{\partial \alpha}{\partial x^j}
- \sum_{j,k} \alpha \beta^j (G_{jk} \hat{T}^{jk} - G_{kj} \hat{T}^{kk})
-  \sum _{j,k} \sigma _{kj} \hat{T}^{jk}  +\alpha\hat{F^0_{LL}},
\label{cmem}
\end{eqnarray}
\begin{eqnarray}
&& \frac{1}{ne} \frac{\partial}{\partial t}
\left ( \frac{\mu h^\ddagger}{ne} \hat{J}^{\dagger i} \right ) =
 - \frac{1}{ne} \left [
\frac{1}{h_1 h_2 h_3} \sum_j  \frac{\partial}{\partial x^j} \left [
\frac{\alpha h_1 h_2 h_3}{h_j} (\hat{K}^{ij}
+  \beta^j \frac{\mu h^\ddagger}{ne} \hat{J}^{\dagger i} ) \right ] \right . \nonumber \\
&& \left . + \frac{2 \mu h^\ddagger}{ne} \frac{1}{h_i}
\frac{\partial \alpha}{\partial x^i} \hat{\rho}_{\rm e}^\dagger
- \sum_j \alpha \left \{ G_{ij} \hat{K}^{ij} - G_{ji} \hat{K}^{jj}
+ \beta^j \frac{\mu h^\ddagger}{ne} \left (G_{ij} \hat{J}^{\dagger i}
- G_{ji} \hat{J}^{\dagger j} \right ) \right \}
+ \sum_j \frac{\mu h^\ddagger}{ne} \sigma_{ji} \hat{J}^{\dagger j}
\right ] \nonumber \\
&& + \alpha \left [
\frac{1}{2ne} \frac{1}{h_i} \frac{\partial}{\partial x^i}
(\Delta \mu p - \Delta p)
+ \left ( \hat{U}^\nu - \frac{\Delta \mu}{ne} \hat{J}^\nu  \right ) \hat{F}_{i\nu}
-\eta [\hat{J}^i - \rho_{\rm e}'(1+\Theta) \hat{U}^i] + \frac{m_- \hat{F}^i_{LL+}-m_+ \hat{F}^i_{LL-}}{\rho e}\right ],
\label{genrelgenohm3+1}
\end{eqnarray}
\begin{eqnarray}
&& \frac{2}{ne} \frac{\partial}{\partial t}
\left [ \frac{\mu h^\ddagger}{ne} \hat{\rho}_{\rm e}^{\dagger}
+\frac{1}{4} (\Delta \mu p - \Delta p) \right ] =
- \frac{1}{ne} \left [
\frac{1}{h_1 h_2 h_3} \sum_j  \frac{\partial}{\partial x^j} \left [
\frac{\alpha h_1 h_2 h_3}{h_j} \frac{\mu h^\ddagger}{ne} (\hat{J}^{\dagger j}
+ 2 \beta^j  \hat{\rho}_{\rm e}^{\dagger} ) \right ] \right . \nonumber \\
&& \left . + \sum_j \frac{\mu h^\ddagger}{ne} \frac{1}{h_j}
\frac{\partial \alpha}{\partial x^j} \hat{J}^{\dagger j}
+\sum_{jk} \alpha \beta^k \left (G_{kj} \hat{K}^{kj}
- G_{jk} \hat{K}^{jj} \right )
+ \sum_{kj} \sigma_{kj} \hat{K}^{kj}
\right ]
\nonumber \\
&& + \alpha \left [
- \frac{1}{2ne} \sum_j \frac{\beta_j}{h_j} \frac{\partial}{\partial x^j}
(\Delta \mu p - \Delta p)
+ \left ( \hat{U}^\nu - \frac{\Delta \mu}{ne} \hat{J}^\nu  \right ) \hat{F}_{\nu 0}
-\eta [\hat{\rho}_{\rm e} - \rho_{\rm e}' (1+\Theta) \hat{\gamma}] + \frac{m_- \hat{F}^0_{LL+}-m_+ \hat{F}^0_{LL-}}{\rho e} \right ],
\label{ohm31z}
\end{eqnarray}
\begin{equation}
\frac{\partial \hat{B}_i}{\partial t} = \frac{- h_i}{h_1 h_2 h_3} \sum _{j,k}
\epsilon ^{ijk} \frac{\partial}{\partial x^j}
\left [ \alpha h_k \left (\hat{E}_k - \sum _{l,m} \epsilon ^{klm} \beta ^l
\hat{B}_m \right )
\right ],
\label{cmfa}
\end{equation}
\begin{equation}
 \sum _{j} \frac{1}{h_1 h_2 h_3} \frac{\partial}{\partial x^j}
\left ( \frac{h_1 h_2 h_3}{h_j} \hat{B}_j
\right ) = 0    ,
\label{glm31}
\end{equation}
\begin{equation}
\hat{\rho}_{\rm e} = \sum _{j} \frac{1}{h_1 h_2 h_3}
\frac{\partial}{\partial x^j} \left (
\frac{h_1 h_2 h_3}{h_j} \hat{E}_j
\right )    ,
\label{dive}
\end{equation}
\begin{eqnarray}
\alpha \left ( \hat{J}^i + \hat{\rho}_{\rm e} \beta ^i \right )
+ \frac{\partial \hat{E}_i}{\partial t} =
\sum _{j,k} \frac{h_i}{h_1 h_2 h_3} \epsilon ^{ijk}
\frac{\partial}{\partial x^j} \left [
\alpha h_k \left ( \hat{B}_k + \sum _{l,m} \epsilon _{klm} \beta ^l \hat{E}_k
\right ) \right ]   ,
\label{cmam}
\end{eqnarray}
where
\begin{eqnarray}
 G_{ij} &=& - \frac{1}{h_i h_j} \frac{\partial h_i}{\partial x^j}\\
 \sigma _{ij} &=& \frac{1}{h_j} \frac{\partial}{\partial x^j}
(\alpha \beta^i)
\end{eqnarray}

\begin{eqnarray}
 \hat{P}^i &=& h \left [h \gamma ^2 \hat{v}^i  +
\frac{\Delta h}{2 n e h} (\hat{U}^i \hat{\rho}_{\rm e} + \hat{J}^i \hat{\gamma})
+ \frac{\mu h^\ddagger}{(ne)^2 h} \hat{J}^i \hat{\rho}_{\rm e}
\right ]
+ (\hat{\bf E} \times \hat{\bf B})_i   ,
\label{gg25}
\\
 \epsilon &=& h \left [ \hat{\gamma} ^2
+  \frac{\Delta h}{neh} \hat{\gamma} \hat{\rho}_{\rm e}
+ \frac{\mu h^\ddagger}{(ne)^2 h} \hat{\rho}_{\rm e}^2 \right ]
-p - \rho \hat{\gamma} +
\frac{\hat{B}^2}{2} + \frac{\hat{E}^2}{2} ,
\label{gg26}
\\
\hat{T}^{ij} &=& p \delta ^{ij} + h \left [
\gamma ^2 \hat{v}^i \hat{v}^j + \frac{\Delta h}{2neh}
(\hat{U}^i \hat{J}^j + \hat{J}^i \hat{U}^j )
+ \frac{\mu h^\ddagger}{(ne)^2 h} \hat{J}^i \hat{J}^j
\right ] \nonumber \\
& +& \left ( \frac{\hat{B}^2}{2} +
\frac{\hat{E}^2}{2} \right) \delta ^{ij} -
\hat{B}_i \hat{B}_j - \hat{E}_i \hat{E}_j   \label{gg27}
\end{eqnarray}

\begin{eqnarray}
\hat{F}^0_{LL-}&=&
\frac{2e^{4}}{3m_-^{2}} \frac{\left  (\rho \gamma - \frac{m_+}{e} {\hat{\rho}}_{\rm e} \right )}{m}\mathbf{\hat{E}} \cdot (\mathbf{\hat{E}}+(\frac{  \rho \gamma \mathbf{\hat{v}} - \frac{m_+}{e} \mathbf{\hat{J}}  }{  \rho \gamma - \frac{m_+}{e} {\hat{\rho}}_{\rm e}  }) \times \mathbf{\hat{B}})
 \nonumber \\
 &-&  \frac{2e^{4}}{3m_-^{2}\displaystyle{\left( 1-{(\frac{  \rho \gamma \mathbf{\hat{v}} - \frac{m_+}{e} \mathbf{\hat{J}}  }{  \rho \gamma - \frac{m_+}{e} {\hat{\rho}}_{\rm e}  })^{2}}\right) }}
 \frac{\left  (\rho \gamma - \frac{m_+}{e} {\hat{\rho}}_{\rm e} \right )}{m}
\left\{ \left( \mathbf{\hat{E}}+(\frac{  \rho \gamma \mathbf{\hat{v}} - \frac{m_+}{e} \mathbf{\hat{J}}  }{  \rho \gamma - \frac{m_+}{e} {\hat{\rho}}_{\rm e}  })\times \mathbf{\hat{B}}\right) ^{2}%
%\right.\nonumber \\
 %- \left.
-\left( (\frac{  \rho \gamma \mathbf{\hat{v}} - \frac{m_+}{e} \mathbf{\hat{J}}  }{  \rho \gamma - \frac{m_+}{e} {\hat{\rho}}_{\rm e}  })\cdot \mathbf{\hat{E}}\right) ^{2}\right\}
\end{eqnarray}

\begin{eqnarray}
\hat{F}^0_{LL+}&=&
\frac{2e^{4}}{3m_+^{2}} \frac{\left  (\rho \gamma + \frac{m_-}{e} {\hat{\rho}}_{\rm e} \right )}{m}\mathbf{\hat{E}} \cdot (\mathbf{\hat{E}}+(\frac{  \rho \gamma \mathbf{\hat{v}} + \frac{m_-}{e} \mathbf{\hat{J}}  }{  \rho \gamma + \frac{m_-}{e} {\hat{\rho}}_{\rm e}  }) \times \mathbf{\hat{B}})
  \nonumber \\
 &-&  \frac{2e^{4}}{3m_+^{2}\displaystyle{\left( 1-{(\frac{  \rho \gamma \mathbf{\hat{v}} + \frac{m_-}{e} \mathbf{\hat{J}}  }{  \rho \gamma + \frac{m_-}{e} {\hat{\rho}}_{\rm e}  })^{2}}\right) }}
\frac{\left  (\rho \gamma + \frac{m_-}{e} {\hat{\rho}}_{\rm e} \right )}{m}
\left\{ \left( \mathbf{\hat{E}}+(\frac{  \rho \gamma \mathbf{\hat{v}} + \frac{m_-}{e} \mathbf{\hat{J}}  }{  \rho \gamma + \frac{m_-}{e} {\hat{\rho}}_{\rm e}  })\times \mathbf{\hat{B}}\right) ^{2}%
%\right.\nonumber \\
 %- \left.
-\left( (\frac{  \rho \gamma \mathbf{\hat{v}} + \frac{m_-}{e} \mathbf{\hat{J}}  }{  \rho \gamma + \frac{m_-}{e} {\hat{\rho}}_{\rm e}  })\cdot \mathbf{\hat{E}}\right) ^{2}\right\}
\end{eqnarray}

\begin{eqnarray}
\hat{F}^i_{LL-} &=&
\frac{2e^{4}}{3m_-^{2}}\frac{\left  (\rho \gamma - \frac{m_+}{e} {\hat{\rho}}_{\rm e} \right )}{m}\left\{ \mathbf{\hat{E}\times \hat{B}}
+
\left( \mathbf{\hat{B}}\times \left( \mathbf{\hat{B}}\times(\frac{  \rho \gamma \mathbf{\hat{v}} - \frac{m_+}{e} \mathbf{\hat{J}}  }{  \rho \gamma - \frac{m_+}{e} {\hat{\rho}}_{\rm e}  })\right)\right)
+
\mathbf{\hat{E}}\left( (\frac{  \rho \gamma \mathbf{\hat{v}} - \frac{m_+}{e} \mathbf{\hat{J}}  }{  \rho \gamma - \frac{m_+}{e} {\hat{\rho}}_{\rm e}  })\cdot \mathbf{\hat{E}}\right) \right\}_i
\nonumber \\
&-&  \frac{2e^{4}}{3m_-^{2}\displaystyle{\left( 1-{(\frac{  \rho \gamma \mathbf{\hat{v}} - \frac{m_+}{e} \mathbf{\hat{J}}  }{  \rho \gamma - \frac{m_+}{e} {\hat{\rho}}_{\rm e}  })^{2}}\right) }}
\frac{\left  (\rho \gamma \mathbf{\hat{v}} - \frac{m_+}{e} \mathbf{\hat{J}} \right )_i}{m}\left\{ \left( \mathbf{\hat{E}}
+
(\frac{ \rho \gamma \mathbf{\hat{v}} - \frac{m_+}{e} \mathbf{\hat{J}}  }{  \rho \gamma - \frac{m_+}{e} {\hat{\rho}}_{\rm e}  })\times \mathbf{\hat{B}}\right) ^{2}-
\left( (\frac{  \rho \gamma \mathbf{\hat{v}} - \frac{m_+}{e} \mathbf{\hat{J}}  }{  \rho \gamma - \frac{m_+}{e} {\hat{\rho}}_{\rm e}  })\cdot \mathbf{\hat{E}}\right) ^{2}\right\}
\end{eqnarray}

\begin{eqnarray}
\hat{F}^i_{LL+} &=&
\frac{2e^{4}}{3m_+^{2}}\frac{\left  (\rho \gamma + \frac{m_-}{e} {\hat{\rho}}_{\rm e} \right )}{m}\left\{ \mathbf{\hat{E}\times \hat{B}}
+
\left( \mathbf{\hat{B}}\times \left( \mathbf{\hat{B}}\times(\frac{  \rho \gamma \mathbf{\hat{v}} + \frac{m_-}{e} \mathbf{\hat{J}}  }{  \rho \gamma + \frac{m_-}{e} {\hat{\rho}}_{\rm e}  })\right)\right)
+
\mathbf{\hat{E}}\left( (\frac{  \rho \gamma \mathbf{\hat{v}} + \frac{m_-}{e} \mathbf{\hat{J}}  }{  \rho \gamma + \frac{m_-}{e} {\hat{\rho}}_{\rm e}  })\cdot \mathbf{\hat{E}}\right) \right\}_i
\nonumber \\
&-&  \frac{2e^{4}}{3m_+^{2}\displaystyle{\left( 1-{(\frac{  \rho \gamma \mathbf{\hat{v}} + \frac{m_-}{e} \mathbf{\hat{J}}  }{  \rho \gamma + \frac{m_-}{e} {\hat{\rho}}_{\rm e}  })^{2}}\right) }}
\frac{\left  (\rho \gamma \mathbf{\hat{v}} + \frac{m_-}{e} \mathbf{\hat{J}} \right )_i}{m}\left\{ \left( \mathbf{\hat{E}}
 +
(\frac{ \rho \gamma \mathbf{\hat{v}} + \frac{m_-}{e} \mathbf{\hat{J}}  }{  \rho \gamma + \frac{m_-}{e} {\hat{\rho}}_{\rm e}  })\times \mathbf{\hat{B}}\right) ^{2}-
\left( (\frac{  \rho \gamma \mathbf{\hat{v}} + \frac{m_-}{e} \mathbf{\hat{J}}  }{  \rho \gamma + \frac{m_-}{e} {\hat{\rho}}_{\rm e}  })\cdot \mathbf{\hat{E}}\right) ^{2}\right\}
\end{eqnarray}

\begin{eqnarray}
\hat{F}^0_{LL}&=&
\frac{2e^{4}}{3m_-^{2}} \frac{\left  (\rho \gamma - \frac{m_+}{e} {\hat{\rho}}_{\rm e} \right )}{m}\mathbf{\hat{E}} \cdot (\mathbf{\hat{E}}+(\frac{  \rho \gamma \mathbf{\hat{v}} - \frac{m_+}{e} \mathbf{\hat{J}}  }{  \rho \gamma - \frac{m_+}{e} {\hat{\rho}}_{\rm e}  }) \times \mathbf{\hat{B}})
 \nonumber \\
 &-&  \frac{2e^{4}}{3m_-^{2}\displaystyle{\left( 1-{(\frac{  \rho \gamma \mathbf{\hat{v}} - \frac{m_+}{e} \mathbf{\hat{J}}  }{  \rho \gamma - \frac{m_+}{e} {\hat{\rho}}_{\rm e}  })^{2}}\right) }}
 \frac{\left  (\rho \gamma - \frac{m_+}{e} {\hat{\rho}}_{\rm e} \right )}{m}
\left\{ \left( \mathbf{\hat{E}}+(\frac{  \rho \gamma \mathbf{\hat{v}} - \frac{m_+}{e} \mathbf{\hat{J}}  }{  \rho \gamma - \frac{m_+}{e} {\hat{\rho}}_{\rm e}  })\times \mathbf{\hat{B}}\right) ^{2}%
%\right.\nonumber \\
 %- \left.
-\left( (\frac{  \rho \gamma \mathbf{\hat{v}} - \frac{m_+}{e} \mathbf{\hat{J}}  }{  \rho \gamma - \frac{m_+}{e} {\hat{\rho}}_{\rm e}  })\cdot \mathbf{\hat{E}}\right) ^{2}\right\}
\nonumber \\
&+&\frac{2e^{4}}{3m_+^{2}} \frac{\left  (\rho \gamma + \frac{m_-}{e} {\hat{\rho}}_{\rm e} \right )}{m}\mathbf{\hat{E}} \cdot (\mathbf{\hat{E}}+(\frac{  \rho \gamma \mathbf{\hat{v}} + \frac{m_-}{e} \mathbf{\hat{J}}  }{  \rho \gamma + \frac{m_-}{e} {\hat{\rho}}_{\rm e}  }) \times \mathbf{\hat{B}})
  \nonumber \\
 &-&  \frac{2e^{4}}{3m_+^{2}\displaystyle{\left( 1-{(\frac{  \rho \gamma \mathbf{\hat{v}} + \frac{m_-}{e} \mathbf{\hat{J}}  }{  \rho \gamma + \frac{m_-}{e} {\hat{\rho}}_{\rm e}  })^{2}}\right) }}
\frac{\left  (\rho \gamma + \frac{m_-}{e} {\hat{\rho}}_{\rm e} \right )}{m}
\left\{ \left( \mathbf{\hat{E}}+(\frac{  \rho \gamma \mathbf{\hat{v}} + \frac{m_-}{e} \mathbf{\hat{J}}  }{  \rho \gamma + \frac{m_-}{e} {\hat{\rho}}_{\rm e}  })\times \mathbf{\hat{B}}\right) ^{2}%
%\right.\nonumber \\
 %- \left.
-\left( (\frac{  \rho \gamma \mathbf{\hat{v}} + \frac{m_-}{e} \mathbf{\hat{J}}  }{  \rho \gamma + \frac{m_-}{e} {\hat{\rho}}_{\rm e}  })\cdot \mathbf{\hat{E}}\right) ^{2}\right\}
\label{gg21}
\end{eqnarray}
\begin{eqnarray}
\hat{F}^i_{LL} &=&
\frac{2e^{4}}{3m_-^{2}}\frac{\left  (\rho \gamma - \frac{m_+}{e} {\hat{\rho}}_{\rm e} \right )}{m}\left\{ \mathbf{\hat{E}\times \hat{B}}
+
\left( \mathbf{\hat{B}}\times \left( \mathbf{\hat{B}}\times(\frac{  \rho \gamma \mathbf{\hat{v}} - \frac{m_+}{e} \mathbf{\hat{J}}  }{  \rho \gamma - \frac{m_+}{e} {\hat{\rho}}_{\rm e}  })\right)\right)
+
\mathbf{\hat{E}}\left( (\frac{  \rho \gamma \mathbf{\hat{v}} - \frac{m_+}{e} \mathbf{\hat{J}}  }{  \rho \gamma - \frac{m_+}{e} {\hat{\rho}}_{\rm e}  })\cdot \mathbf{\hat{E}}\right) \right\}_i
\nonumber \\
&-&  \frac{2e^{4}}{3m_-^{2}\displaystyle{\left( 1-{(\frac{  \rho \gamma \mathbf{\hat{v}} - \frac{m_+}{e} \mathbf{\hat{J}}  }{  \rho \gamma - \frac{m_+}{e} {\hat{\rho}}_{\rm e}  })^{2}}\right) }}
\frac{\left  (\rho \gamma \mathbf{\hat{v}} - \frac{m_+}{e} \mathbf{\hat{J}} \right )_i}{m}\left\{ \left( \mathbf{\hat{E}}
+
(\frac{ \rho \gamma \mathbf{\hat{v}} - \frac{m_+}{e} \mathbf{\hat{J}}  }{  \rho \gamma - \frac{m_+}{e} {\hat{\rho}}_{\rm e}  })\times \mathbf{\hat{B}}\right) ^{2}-
\left( (\frac{  \rho \gamma \mathbf{\hat{v}} - \frac{m_+}{e} \mathbf{\hat{J}}  }{  \rho \gamma - \frac{m_+}{e} {\hat{\rho}}_{\rm e}  })\cdot \mathbf{\hat{E}}\right) ^{2}\right\}
\nonumber \\
&+&\frac{2e^{4}}{3m_+^{2}}\frac{\left  (\rho \gamma + \frac{m_-}{e} {\hat{\rho}}_{\rm e} \right )}{m}\left\{ \mathbf{\hat{E}\times \hat{B}}
+
\left( \mathbf{\hat{B}}\times \left( \mathbf{\hat{B}}\times(\frac{  \rho \gamma \mathbf{\hat{v}} + \frac{m_-}{e} \mathbf{\hat{J}}  }{  \rho \gamma + \frac{m_-}{e} {\hat{\rho}}_{\rm e}  })\right)\right)
+
\mathbf{\hat{E}}\left( (\frac{  \rho \gamma \mathbf{\hat{v}} + \frac{m_-}{e} \mathbf{\hat{J}}  }{  \rho \gamma + \frac{m_-}{e} {\hat{\rho}}_{\rm e}  })\cdot \mathbf{\hat{E}}\right) \right\}_i
\nonumber \\
&-&  \frac{2e^{4}}{3m_+^{2}\displaystyle{\left( 1-{(\frac{  \rho \gamma \mathbf{\hat{v}} + \frac{m_-}{e} \mathbf{\hat{J}}  }{  \rho \gamma + \frac{m_-}{e} {\hat{\rho}}_{\rm e}  })^{2}}\right) }}
\frac{\left  (\rho \gamma \mathbf{\hat{v}} + \frac{m_-}{e} \mathbf{\hat{J}} \right )_i}{m}\left\{ \left( \mathbf{\hat{E}}
 +
(\frac{ \rho \gamma \mathbf{\hat{v}} + \frac{m_-}{e} \mathbf{\hat{J}}  }{  \rho \gamma + \frac{m_-}{e} {\hat{\rho}}_{\rm e}  })\times \mathbf{\hat{B}}\right) ^{2}-
\left( (\frac{  \rho \gamma \mathbf{\hat{v}} + \frac{m_-}{e} \mathbf{\hat{J}}  }{  \rho \gamma + \frac{m_-}{e} {\hat{\rho}}_{\rm e}  })\cdot \mathbf{\hat{E}}\right) ^{2}\right\}
\label{gg22}
\end{eqnarray}
Equation (\ref{gg21}) to (\ref{gg22}) could be changed to the following with $m=m_++ m_-$, $\mu = m_+ m_-/m^2$, $\Delta \mu = (m_+ -m_-)/m$

\begin{eqnarray}
\hat{F}^0_{LL}&=&\frac{2e^4}{3m^3\mu^2}\mathbf{\hat{E}}\cdot[\rho\gamma(\mathbf{\hat{E}}+\mathbf{\hat{v}}\times\mathbf{\hat{B}})(1-2\mu)-\frac{\hat{\rho}_e\mathbf{\hat{E}}+\mathbf{\hat{J}}\times\mathbf{\hat{B}}}{e}(\Delta \mu m(1-\mu))]
\nonumber \\
&-&\frac{2e^4}{3m^3\mu^2}\{(a_1 b_1+a_2 b_2m^4\mu^2-a_3 b_3m^2\mu)[\rho\gamma(1-2\mu)-\frac{\hat{\rho}_e}{e}(\Delta\mu m(1-\mu))]
\nonumber \\
&+&(a_3b_2m^2\mu-a_1b_3)[\rho\gamma(\Delta\mu m(1-\mu))-\frac{\hat{\rho}_e}{e}((m-2m\mu)^2-2m^2\mu^2)]
\nonumber \\
&+&(a_3b_1\mu-a_2b_3m^2\mu^2)[\rho\gamma\Delta \mu m - \frac{\hat{\rho}_e}{e}(m^2-2m^2\mu)]
\nonumber \\
&+&a_1b_2[\rho\gamma((m-2m\mu)^2-2m^2\mu^2)-\frac{\hat{\rho}_e}{e}(m\Delta \mu((m-2m\mu)^2-2m^2\mu^2+\mu(m^2-2m^2\mu)+m^2\mu^2))]
\nonumber \\
&+&a_2b_1m^2\mu^2[2\rho\gamma-\frac{\hat{\rho}_e}{e}m\Delta \mu]\}/
[a_1^2+a_1a_2(m^2-2m^2\mu)-a_1a_3m\Delta \mu +a_2a_3m^3\mu\Delta\mu+a_2^2m^4\mu^2-a_3^2m^2\mu]\\ \label{gg23}
\hat{F}^i_{LL}&=&\frac{2e^4}{3m}\{\frac{(1-2\mu)\rho\gamma}{m^2\mu^2}[\mathbf{\hat{E}}\times\mathbf{\hat{B}}+\mathbf{\hat{B}}\times(\mathbf{\hat{B}}\times\mathbf{\hat{v}})+(\mathbf{\hat{v}}\cdot\mathbf{\hat{E}})\mathbf{\hat{E}}]
-\frac{\Delta\mu(1-\mu)}{m\mu^2e}[\hat{\rho}_e(\mathbf{\hat{E}}\times\mathbf{\hat{B}})+\mathbf{\hat{B}}\times(\mathbf{\hat{B}}\times\mathbf{\hat{J}})+(\mathbf{\hat{J}}\cdot\mathbf{\hat{E}})\mathbf{\hat{E}}]\}_i
\nonumber \\
 &-&\frac{2e^4}{3m^3\mu^2}\{(a_1 b_1+a_2 b_2m^4\mu^2-a_3 b_3m^2\mu)[\rho\gamma\mathbf{\hat{v}}(1-2\mu)-\frac{\mathbf{\hat{J}}}{e}(\Delta\mu m(1-\mu))]_i
\nonumber \\
&+&(a_3b_2m^2\mu-a_1b_3)[\rho\gamma\mathbf{\hat{v}}(\Delta\mu m(1-\mu))-\frac{\mathbf{\hat{J}}}{e}((m-2m\mu)^2-2m^2\mu^2)]_i
\nonumber \\
&+&(a_3b_1\mu-a_2b_3m^2\mu^2)[\rho\gamma\mathbf{\hat{v}}\Delta \mu m - \frac{\mathbf{\hat{J}}}{e}(m^2-2m^2\mu)]_i
\nonumber \\
&+&a_1b_2[\rho\gamma\mathbf{\hat{v}}((m-2m\mu)^2-2m^2\mu^2)-\frac{\mathbf{\hat{J}}}{e}(m\Delta \mu((m-2m\mu)^2-2m^2\mu^2+\mu(m^2-2m^2\mu)+m^2\mu^2))]_i
\nonumber \\
&+&a_2b_1m^2\mu^2[2\rho\gamma\mathbf{\hat{v}}-\frac{\mathbf{\hat{J}}}{e}m\Delta \mu]\}/
[a_1^2+a_1a_2(m^2-2m^2\mu)-a_1a_3m\Delta \mu +a_2a_3m^3\mu\Delta\mu+a_2^2m^4\mu^2-a_3^2m^2\mu]_i \label{gg24}
\end{eqnarray}

where

\begin{eqnarray*}
a_1&=&\rho^2\gamma^2-\rho^2\gamma^2 \hat{v}^2\\
a_2&=&\frac{\hat{\rho}_e^2-\hat{J}^2}{e^2}\\
a_3&=&\frac{2\rho\gamma\hat{\rho}_e-2\rho\gamma\mathbf{\hat{v}}\cdot\mathbf{\hat{J}}}{e}\\
b_1&=&\rho^2\gamma^2(\mathbf{\hat{E}}+\mathbf{\hat{v}}\times\mathbf{\hat{B}})^2-\rho^2\gamma^2(\mathbf{\hat{v}}\cdot\mathbf{\hat{E}})^2\\
b_2&=&\frac{(\hat{\rho}_e\mathbf{\hat{E}}+\mathbf{\hat{J}}\times\mathbf{\hat{B}})^2-(\mathbf{\hat{J}}\cdot\mathbf{\hat{E}})^2}{e^2}\\
b_3&=&\frac{2\rho\gamma(\mathbf{\hat{E}}+\mathbf{\hat{v}}\times\mathbf{\hat{B}})\cdot(\hat{\rho}_e\mathbf{\hat{E}}+\mathbf{\hat{J}}\times\mathbf{\hat{B}})-2\rho\gamma(\mathbf{\hat{v}}\cdot\mathbf{\hat{E}})(\mathbf{\hat{J}}\cdot\mathbf{\hat{E}})}{e}
\end{eqnarray*}

We find that formulas (\ref{gg25})-(\ref{gg24}) measured in the ZAMO frame are similar to that of special relativistic case.

\section{A Simple Application to Reconnection around the Rotating Black Hole}

\begin{figure*}[t!]
  \begin{center}
    \begin{tabular}{cc}
      \includegraphics[width=0.5\textwidth]{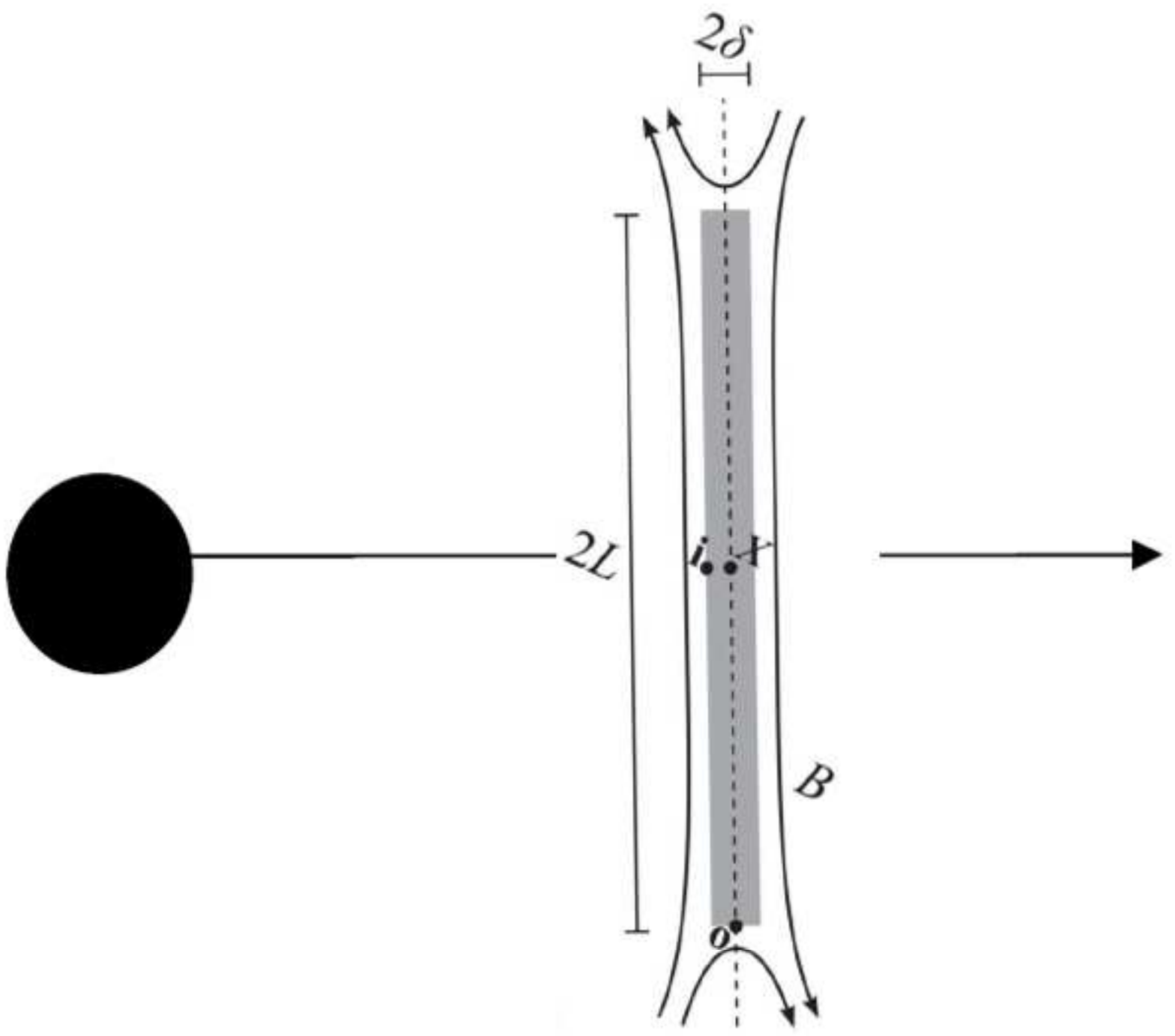}
      \includegraphics[width=0.5\textwidth]{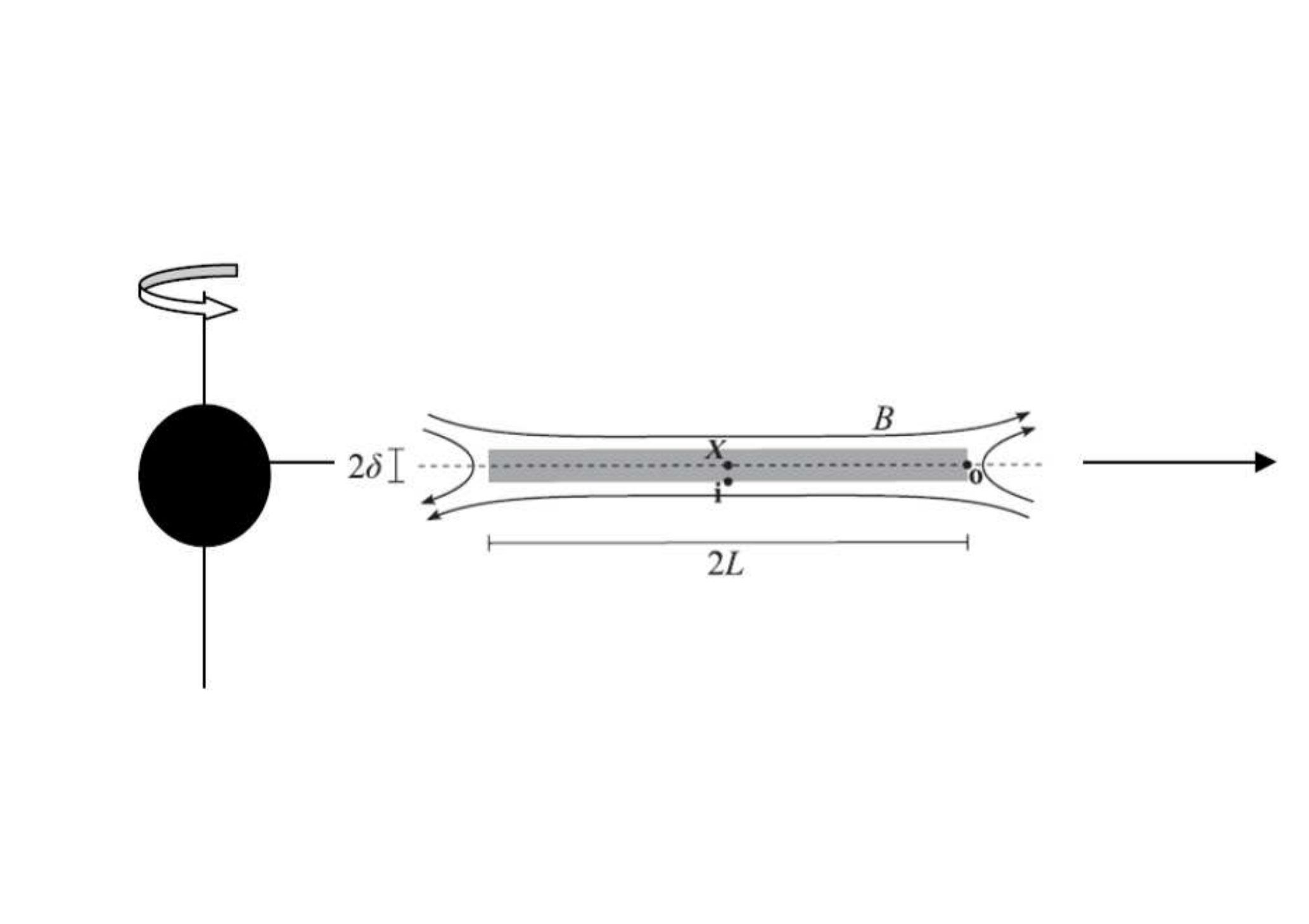}
    \end{tabular}
    \end{center}
     \caption{Left panel is the diagram of a reconnection layer in the azimuthal direction and the rotation of the black hole is clockwise. Right panel is a diagram of a reconnection layer in the radial direction.}
     \label{fig:figure1}
\end{figure*}

Next we analyze magnetic reconnection near black hole with Landau-Lifshitz radiation reaction force shown in Figure.~\ref{fig:figure1}. We consider the spacetime to be a Kerr metric represented by $(t,x^1,x^2,x^3)=(t,r,\theta,\phi)$ and
\begin{eqnarray}
h_0 = (1 - 2{r_g}r/{\Sigma})^{1/2}  \, , \quad h_1 = ({\Sigma}/{\Delta })^{1/2} \, , \nonumber \\
h_2 = \Sigma^{1/2} \, , \quad h_3 = (A/{\Sigma})^{1/2} \sin \theta \, , \; \; \quad \\
\omega_1 = \omega_2 = 0 \, , \quad \omega_3 = {{2r_g^2ar}}/{\Sigma} \, . \; \; \qquad  \nonumber
\end{eqnarray}
where
\begin{eqnarray}
\Sigma  = {r^2} + {\left( {a{r_g}} \right)^2}{\cos ^2}\theta\\
 \Delta  = {r^2} - 2{r_g}r + {\left( {a{r_g}} \right)^2}\\
 A = \big[ {{r^2} + {{\left( {a{r_g}} \right)}^2}} \big]^2 - \Delta {\left( {a{r_g}} \right)^2}{\sin ^2}\theta
\end{eqnarray}
Then we get $\alpha  = (\Delta \Sigma /A)^{1/2}$, $\beta^\phi=h_3\omega_3/\alpha$, $\beta^r=0$ and $\beta^\theta=0$

When we assume the plasma is proton-electron fluid, then equation (\ref{gg28}) reduces to
\begin{eqnarray}
\nabla_\nu  \left [
h \left (U^\mu U^\nu \right ) \right ]
&=& -\nabla^\mu p + J^\nu {F^\mu}_\nu
 \nonumber \\
& - & \frac{2n_- e^{4}}{3m^{2}_-}\left[ F^{\mu
 \lambda }F_{\nu \lambda }u^{\nu }_-
 - \left( F_{\nu \lambda }u^{\lambda }_-\right)
 \left( F^{\nu \kappa }u_{- \kappa }\right) u^{\mu }_-\right]
 \label{gg29}
\end{eqnarray}
where we neglect the radiation reaction force acting on the proton due to its large mass

Equation (\ref{gg29}) could be changed to the following form in the ZAMO frame in the azimuthal direction along the
neutral line \citep{775}
\begin{equation}\label{mo22}
\frac{1}{\alpha h_1h_2}\frac{\partial}{\partial \phi} \left[\alpha h_1h_2 h \gamma^2 \hat v^\phi\left(\hat v^\phi+\beta^\phi\right)\right] =  -h_3 {\hat J}^\theta {\hat B}_r - \frac{\partial p}{\partial \phi}+h_3\hat{F^\phi_{LL}}
\end{equation}

In the radial direction along the
neutral line, equation (\ref{gg29}) could be reduced to \citep{775}
\begin{eqnarray}
\frac{\partial}{\partial r}\left[\alpha h_2h_3 h \gamma^2(\hat v^r)^2\right]+h_2h_3\frac{\partial\alpha}{\partial r}h\gamma^2=
-\alpha {h_2}{h_3}\left( {\frac{{\partial p}}{{\partial r}} + h_1 {{\hat J}^\phi }{{\hat B}_\theta }} -h_1\hat{F^r_{LL}}\right) \, .
\end{eqnarray}

According to the result in \cite{775}, we get ${\hat J}^\theta$ and ${\hat B}_r$ at the outflow point represented by "o" in Figure.~\ref{fig:figure1}. Then the ratio of radiation reaction force to Lorentz force is $\frac{2}{3} nr_e^2\eta c$ (the detailed derivation for the reconnection layer in the azimuthal direction can be found in Section 3.1 and the reconnection layer in the radial direction could be achieved in the same way). When we set $\eta=10^{-2}$, the value of $\frac{2}{3} nr_e^2\eta c$ is $\sim 10^{-18}n$. If we set the density of the plasma to be $\rho=10^{-6}g/cm^3$, the number density is $10^{18}$. Then we note that the radiation reaction force is comparable to Lorentz force and this is the maximum Lorentz force allowed in the physical system in this situation. With the same approximation and calculation used in \cite{775}, we find that the reconnection rate in the azimuthal direction with the radiation reaction can be larger than that without the radiation reaction according to equation (\ref{mo22}) from above and equation (9) in \cite{775}. The outflow during the process of magnetic reconnection is mainly along the neutral line. The results from the radial direction can be achieved using the same method and is similar to the above.

\subsection{Derivation for the reconnection layer in the azimuthal direction}
The Landau-Lifshitz radiation reaction force from the right side of equation (\ref{gg22}) acting on the proton-electron fluid element is (the radiation reaction force acting on the proton is neglected due to $m_p \gg m_e$. Here, we conduct the derivation without setting $c=1$)
\begin{eqnarray*}
\mathbf{\hat F_{LL}} &= &
\frac{2e^{4}}{3m_-^{2}c^4}\frac{\left  (\rho \gamma - \frac{m_+}{e} {\hat{\rho}}_{\rm e} \right )}{m}\left\{ \mathbf{\hat{E}\times \hat{B}}
 +
\frac{1}{c}\left( \mathbf{\hat{B}}\times \left( \mathbf{\hat{B}}\times(\frac{  \rho \gamma \mathbf{\hat{v}} - \frac{m_+}{e} \mathbf{\hat{J}}  }{  \rho \gamma - \frac{m_+}{e} {\hat{\rho}}_{\rm e}  })\right)\right)
+
\frac{1}{c}\mathbf{\hat{E}}\left( (\frac{  \rho \gamma \mathbf{\hat{v}} - \frac{m_+}{e} \mathbf{\hat{J}}  }{  \rho \gamma - \frac{m_+}{e} {\hat{\rho}}_{\rm e}  })\cdot \mathbf{\hat{E}}\right) \right\}
\nonumber \\
&-&  \frac{2e^{4}}{3m_-^{2}c^5\displaystyle{\left( 1-\frac{1}{c^2}{(\frac{  \rho \gamma \mathbf{v} - \frac{m_+}{e} \mathbf{\hat{J}}  }{  \rho \gamma - \frac{m_+}{e} {\hat{\rho}}_{\rm e}  })^{2}}\right) }}
\frac{\left  (\rho \gamma \mathbf{\hat{v}} - \frac{m_+}{e} \mathbf{\hat{J}} \right )}{m}\left\{ \left( \mathbf{\hat{E}}
 +
(\frac{1}{c}\frac{ \rho \gamma \mathbf{\hat{v}} - \frac{m_+}{e} \mathbf{\hat{J}}  }{  \rho \gamma - \frac{m_+}{e} {\hat{\rho}}_{\rm e}  })\times \mathbf{\hat{B}}\right) ^{2}-
\frac{1}{c^2}\left( (\frac{  \rho \gamma \mathbf{\hat{v}} - \frac{m_+}{e} \mathbf{\hat{J}}  }{  \rho \gamma - \frac{m_+}{e} {\hat{\rho}}_{\rm e}  })\cdot \mathbf{\hat{E}}\right) ^{2}\right\} \label{eq151}\\
\end{eqnarray*}

When we assume the plasma is neutral, ${\hat{\rho}}_{\rm e}=0$, then the above equation changes to
\begin{eqnarray*}
\mathbf{\hat F_{LL}}& =&
\frac{2e^{4}}{3m_-^{2}c^4}\frac{\rho \gamma }{m}\left\{ \mathbf{\hat{E}\times \hat{B}}
 +
\frac{1}{c}\left( \mathbf{\hat{B}}\times \left( \mathbf{\hat{B}}\times\frac{  \rho \gamma \mathbf{\hat{v}} - \frac{m_+}{e} \mathbf{\hat{J}}  }{  \rho \gamma   }\right)\right)
+
\frac{1}{c}\mathbf{\hat{E}}\left( \frac{  \rho \gamma \mathbf{\hat{v}} - \frac{m_+}{e} \mathbf{\hat{J}}  }{  \rho \gamma   }\cdot \mathbf{\hat{E}}\right) \right\}
\nonumber \\
&-&  \frac{2e^{4}}{3m_-^{2}c^5\displaystyle{\left( 1-\frac{1}{c^2}{(\frac{  \rho \gamma \mathbf{\hat{v}} - \frac{m_+}{e} \mathbf{\hat{J}}  }{  \rho \gamma   })^{2}}\right) }}
\frac{\left  (\rho \gamma \mathbf{\hat{v}} - \frac{m_+}{e} \mathbf{\hat{J}} \right )}{m}\left\{ \left( \mathbf{\hat{E}}
 +
\frac{1}{c}\frac{ \rho \gamma \mathbf{\hat{v}} - \frac{m_+}{e} \mathbf{\hat{J}}  }{  \rho \gamma   }\times \mathbf{\hat{B}}\right) ^{2}-
\frac{1}{c^2}\left( \frac{  \rho \gamma \mathbf{\hat{v}} - \frac{m_+}{e} \mathbf{\hat{J}}  }{  \rho \gamma   }\cdot \mathbf{\hat{E}}\right) ^{2}\right\} \label{eq151}\\
\end{eqnarray*}

In resistive relativistic magnetohydrodynamics, we assume $\mathbf{\hat{E}}=\eta\mathbf{\hat{J}}-\frac{1}{c}\mathbf{\hat{v}} \times \mathbf{\hat{B}}$ for simplicity. According to the assumption in \cite{775}, $\hat v^r$ vanishes, $\hat B^\phi =0$, and $\hat J^\phi \approx 0 \approx \hat J^r$ at the neutral line and $\hat v^\theta \approx 0$, $\hat B^\theta \approx 0$ and $\partial_\theta \approx 0$ everywhere. The result shows that the outflow velocity is mildly relativistic meaning $\gamma\approx 1$. For convenience of the following derivation, we set $\hat{v}_\phi \approx c$. Substituting these into the above equation, we get the following in the $\phi$ direction

\begin{eqnarray*}
\hat{F}^\phi_{LL}  &=&  -\frac{2}{3} nr_e^2\eta\hat{J}^\theta\hat{B}_r
-\frac{2}{3} n r_e^2[(\eta\hat{J}^\theta)^2+(\frac{\hat{J}^\theta\hat{B}_r}{cne})^2-(\frac{\hat{J}^\theta(\eta\hat{J}^\theta-\frac{\hat{v}_\phi}{c}\hat{B}_r)}{cne})^2]
\nonumber \\
&=&-\frac{2}{3} nr_e^2\eta\hat{J}^\theta\hat{B}_r
-\frac{2}{3} n r_e^2[(\eta\hat{J}^\theta)^2+\frac{[\hat{J}^\theta\hat{B}_r+\hat{J}^\theta(\eta\hat{J}^\theta-\frac{\hat{v}_\phi}{c}\hat{B}_r)][\hat{J}^\theta\hat{B}_r-\hat{J}^\theta(\eta\hat{J}^\theta-\frac{\hat{v}_\phi}{c}\hat{B}_r)]}{(cne)^2}]
\nonumber \\
&=&-\frac{2}{3} nr_e^2\eta\hat{J}^\theta\hat{B}_r
-\frac{2}{3} n r_e^2[(\eta\hat{J}^\theta)^2+\frac{\eta\hat{J}^{\theta 2}(2\hat{J}^\theta\hat{B}_r-\eta\hat{J}^{\theta 2})}{(cne)^2}]
\end{eqnarray*}

Then the ratio of radiation reaction force to the Lorentz force is
\begin{eqnarray*}
\frac{\hat{F}^\phi_{LL}}{\hat{F}^\phi_{L}}&=&\frac{2}{3} nr_e^2\eta c+\frac{2}{3} n r_e^2 \eta c \frac{\eta\hat{J}^\theta}{\hat{B}_r}+\frac{2}{3} n r_e^2 \eta c \frac{2\hat{J}^{\theta 2}\hat{B}_r-\eta\hat{J}^{\theta 3}}{(cne)^2\hat{B}_r}
\nonumber \\
&=&\frac{2}{3} nr_e^2\eta c+\frac{2}{3} n r_e^2 \eta c \frac{\eta\hat{J}^\theta}{\hat{B}_r}+\frac{2}{3} n r_e^2 \eta c[2(\frac{\hat{v}_{ie}}{c})^2-\frac{\eta\hat{J}^\theta}{\hat{B}_r}(\frac{\hat{v}_{ie}}{c})^2]
\nonumber \\
&=&\frac{2}{3} nr_e^2\eta c[1+\frac{\eta\hat{J}^\theta}{\hat{B}_r}+(2(\frac{\hat{v}_{ie}}{c})^2-\frac{\eta\hat{J}^\theta}{\hat{B}_r}(\frac{\hat{v}_{ie}}{c})^2)]
\nonumber \\
& \approx & \frac{2}{3} nr_e^2\eta c
\end{eqnarray*}
where we assume that the small value of $\eta$ is small enough, and that the relative velocity of proton and electron in the $\theta$ direction, $\hat v_{ie}$, is far less than the speed of light.

Then the radiation reaction force in the azimuthal direction could be written as
\begin{equation*}
\hat{F}^\phi_{LL} \approx -\frac{2}{3} nr_e^2\eta \hat{J}^\theta\hat{B}_r
\end{equation*}

\section{Discussion}
We have derived the GRMHD equations of one-fluid by using the two-fluid approximation of plasma made of positively and negatively charged particles into which the Landau-Lifshitz radiation reaction force is incorporated in curved space, providing a detailed self-consistent expression of the Landau-Lifshitz radiation reaction force felt by the charged particles in the plasma. These derived results could apply to two-fluid of the arbitrary mass ratio of a positively charged particle to a negatively charged particle, like positron-electron or proton-electron plasma. The GRMHDs containing the Landau-Lifshitz radiation reaction force is a natural generalization of modern GRMHDs and thus could be applied to many physical processes in astrophysics for simulation, like the accretion onto a newly born stellar mass black hole in a gamma-ray burst and the inner accretion region around neutron star where the motion of plasma is relativistic, the background magnetic field is sufficiently large and the curved space is obvious. Because of the small value of radiation reaction force compared to the Lorentz force in most of previous studies, the radiation reaction is not considered to be a major contribution to the dynamics of plasma and thus is usually neglected. When accounting for the radiation reaction force in curved space by adopting the general relativistic particle-in-cell method, \cite{5} showed that the current layers become thinner in contrast to that without a radiation reaction in the pulsar magnetosphere. Thus, in the extreme astrophysical process, we expect that there may exist some distinctive phenomenon. Let us take the accretion around a newly born magnetized black hole in a gamma-ray burst for instance. First we investigate the motion of a test electron moving around the magnetized black hole in a stable orbit perpendicular to the large magnetic field. We find that the kinetic energy of the test electron decreases along with motion due to the radiation reaction force.
According to the results in \cite{49}, the radiation reaction will make the electron fall into the black hole when the direction of the Lorentz force is toward the black hole, while when Lorentz force is outward of the black hole, the orbit of the electron could remain bounded and oscillations are decaying. When considering an accretion disk around such a black hole, we may expect that the accretion rate of material flowing into the black hole with a radiation reaction is different from that without a radiation reaction; thus, the radiation reaction force can affect the accretion of charged particles in an accretion disk around the black hole significantly. Numerical simulation applied to high energy astrophysics with equations derived in this work will be performed in the future.

\section*{Acknowledgements}
%We are very grateful to the anonymous referee for her/his instructive comments which improved the content of the paper.
We are very grateful to the anonymous referee for the instructive comments, which improved the content of the manuscript. This work is supported by the National Key Research and Development Program of China (No. 2017YFA0402703). This work
has been supported by the National Science Foundations of China
(Nos. 11373024, 11233003 and 11873032).

%--------------------------------------------------%

%--------------------------------------------------%

\bibliographystyle{apj}

\end{document}